\documentclass[amsmath,amssymb]{article}
\usepackage{graphicx}
\usepackage{lipsum}
\usepackage{amsmath}

\begin{document}
\title{Observation of elliptical rings in Type-I spontaneous parametric
  down-conversion}
\author{Hannah Guilbert, Yu-Po Wong, Daniel J. Gauthier}
\date{\today}

\maketitle
\begin{abstract}
We investigate the transverse spatial profile of down-converted light produced by noncollinear, degenerate, Type-I spontaneous parametric down-conversion in 
two types of nonlinear crystals. We find that the pattern produced by one crystal, beta barium borate (BBO), produces a
circular down-conversion pattern while the other crystal, bismuth triborate (BiBO) produces an elliptical pattern. We show this difference is due to the angle-independent refractive index experienced by the daughter photons in BBO, while they experience an angle-dependent refractive index in BiBO. We image the transverse
spatial profile of the generated light to determine the eccentricity produced by each crystal and develop a model to explain our observation. Among other things, this model predicts that there is a wavelength for which the eccentricity from
BiBO is nearly zero. Finally, we discuss how the elliptical ring pattern produced in BiBO potentially affects polarization entanglement for experimental setups that collect biphotons around the entire down-conversion ring. We show that the quality of polarization entanglement as measured by the overlap integral of the spectrum of the two rings, can remain high ($>99.4\%$) around the entire ring at the expense of decreased biphoton rate. 
\end{abstract}



\section{Introduction}
The nonlinear optical process of spontaneous parametric down-conversion (SPDC) is a popular method for producing correlated photon pairs, also referred to as biphotons or daughter photons. A major advantage of this process is the 
ability to quantum mechanically entangle the daughter photons in various degrees-of-freedom including polarization, time, frequency, space, and momentum, for example. 
The entanglement quality can be very high, making it an attractive source for applications in quantum communication, such as quantum key distribution (QKD) \cite{BB84, Ekert, Serigenko}, as well as experiments in fundamental quantum information science \cite{Zeilenger,bradprl}.

In the SPDC process, a pump photon ($p$) enters a nonlinear optical crystal and is annihilated, producing two daughter photons, typically called 
signal ($s$) and idler ($i$). Energy conservation dictates that $\omega_{p} = \omega_{s}+\omega_{i}$. To be an efficient process, 
the photons must also abide by momentum conservation, a condition referred to as phase matching. In noncollinear SPDC, the daughter photons are emitted into angles on either side 
of the pump direction
in order to satisfy this phase matching relation. In degenerate SPDC where $\omega_{s}=\omega_{i}$, this emission occurs in opposite pairs of directions around the pump beam forming a ring, called the down-conversion ring, in a plane transverse to the pump.
 
In many experiments aimed at creating entangled photons pairs, the polarization degree-of-freedom is used because the measurement process is straightforward and very high
purity can be achieved. Because current QKD schemes aim at maximizing photon rates, methods of creating high brightness polarization-entangled sources have been investigated \cite{Kim,Wong,Kwiat95, Kwiat99, Kwiat05, Kwiat09}. A potential solution involves using Type-I phase matching, where the
 two daughter photons have the same polarization that is opposite to the pump. Recently, Rangarajan \textit{et al.} \cite{Kwiat09} showed that high-brightness polarization entanglement can be achieved in
 Type-I phase matching when two thin crystals are placed together with their optic axes rotated $90^{\circ}$ with respect to each other \cite{Kwiat05,Kwiat09}. Originally, the process was demonstrated in 
 beta barium borate (BBO, a uniaxial crystal) \cite{Kwiat99, Kwiat05}, but later shown for bismuth triborate, (BiBO, a biaxial crystal) \cite{Kwiat09} which is quite promising for high-brighness applications because of its higher nonlinear coefficient  \cite{Hellwig,Ghotbi}.
 
\begin{figure}
\begin{center}
 \includegraphics[scale=0.4]{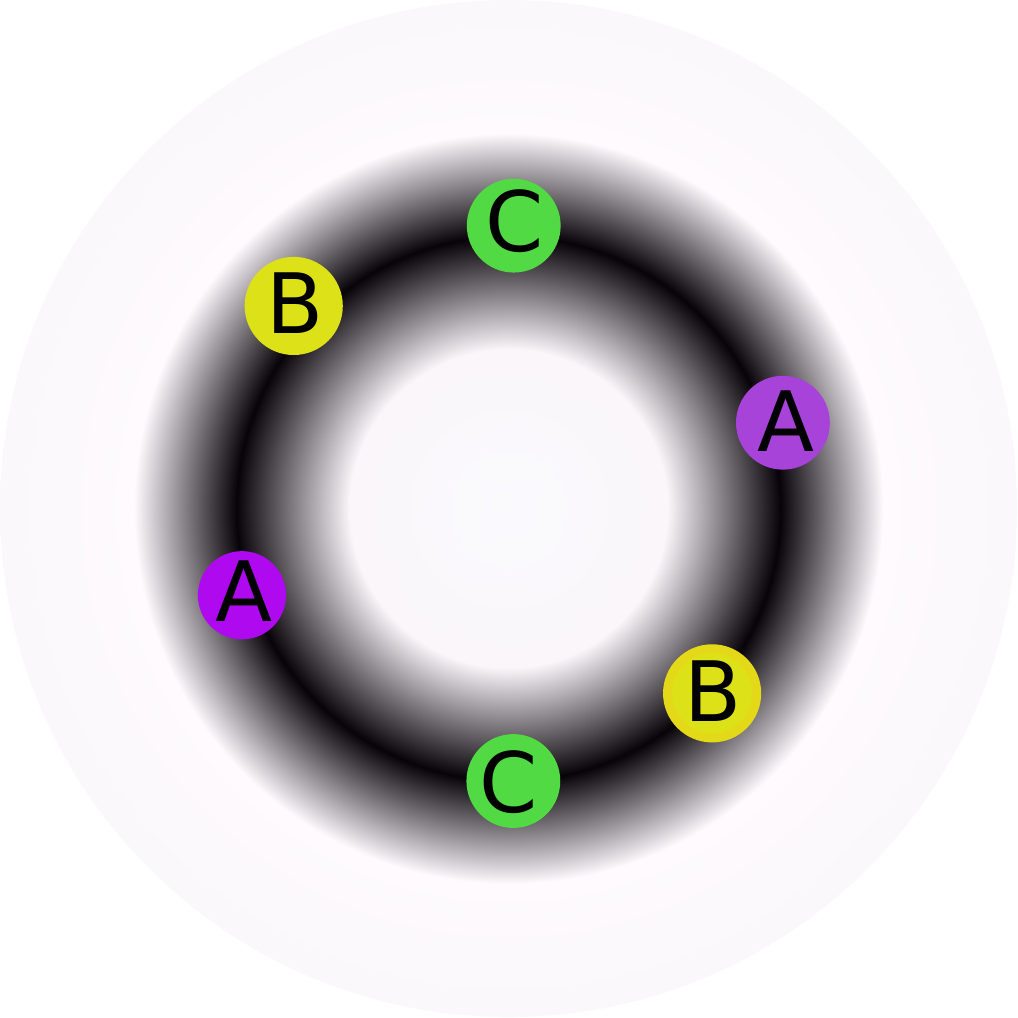}
\end{center}
\caption{The dark circle illustrates the down-conversion ring. Points A, B, and C represent pairs of daughter photons at different azimuthal angles around the ring.
 To increase the total count rate, one should collect from multiple points around the ring. This idea is called spatial multiplexing.}
\label{multiplex}
\end{figure}

 Each crystal produces the same polarization but there is essentially no which-path information regarding the crystal in which the photons are generated. This process can give rise to polarization entanglement around the entire ring, if the rings overlap everywhere. This enables higher achievable fluxes by allowing for collection of biphotons at multiple pairs of points around the down-conversion ring,
known as spatial multiplexing and illustrated in Fig. \ref{multiplex}. 

To achieve entanglement around the whole ring, the collected light must be indistinguishable. This only occurs when 
the emmission patterns are completely circular because any eccentricity of the ring from the first crystal will have its major axis perpendicular to the major axis from the ring produced in the second crystal. This leads to reduced spatial overlap and reduced entanglement.  

Although BiBO is a promising crystal for high-brightness SPDC applications, we find that the emission pattern produced from a Type-I interaction with BiBO is elliptical. Eccentricity 
introduces a possible way of distinguishing photons that were born in the first crystal from those born in the second, thereby potentially reducing the entanglement quality around the
 entire ring.

In the next section, we develop a formalism for predicting theoretically the emission pattern for noncollinear degenerate Type-I SPDC in BiBO and discuss the physical reason
for elliptical emission patterns. In Sec. 3, we discuss our experimental setup and present 
our data. In Sec. 4, we discuss a theoretical model for predicting the eccentricity in a biaxial crystal. We show that this model agrees with our experimental data and that there
is a wavelength that minimizes the eccentricity for a given set of emission angles. In Sec. 5, we discuss 
the repercussions of elliptical rings on the entanglement quality and brightness and further show how the spectrum of the single photons changes as a result of the asymmetry. 

\section{Phase Matching for Type-I SPDC in BBO and BiBO}
\begin{figure}[!h]
\begin{center}
 \includegraphics[scale=0.32]{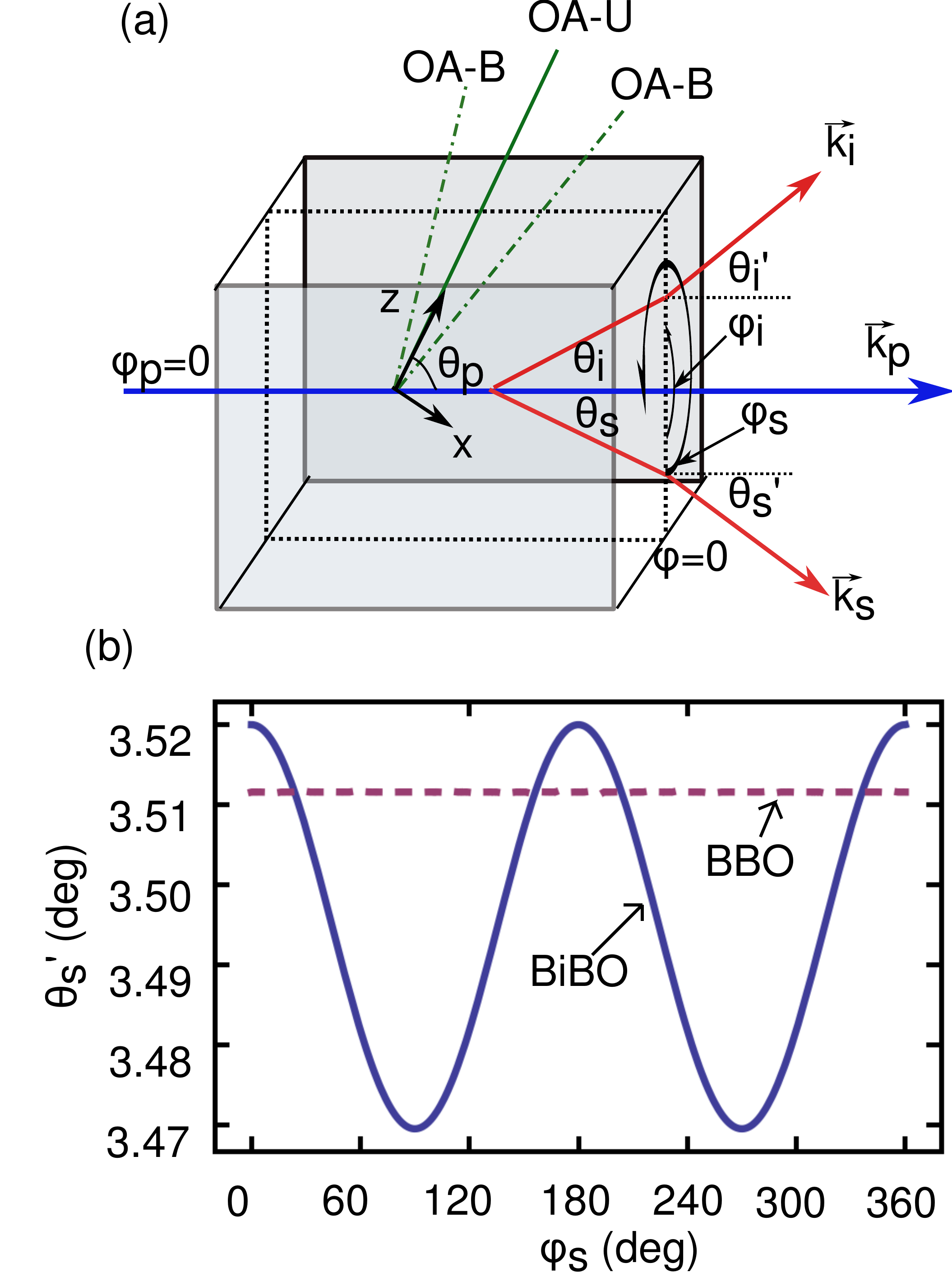}
\end{center}
\caption{(a) Crystal geometry for BBO and BiBO crystals. 
The pump makes angles ($\theta_{p},\phi_{p}$) with respect to the crystal optic axis. Here, $\phi_{p}=0$. The signal (idler) photons are emitted at local angles
 ($\theta_{s(i)},\phi_{s(i)}$) with respect to $\vec{k}_{p}$. Refraction at the interface of the crystal results in an exterior angle of
 $\theta_{s(i)}'$ outside the crystal. (b) The exterior emission angle versus its local azimuthal angle for BiBO (blue, solid line) 
and BBO (maroon, dashed line). For BBO, we use a pump cut angle of $\theta_{p}=29.392^{\circ}$ while for BiBO we use $\phi_{p}=90^{\circ},\theta_{p}=151.563^{\circ}$.
 Both curves are calculated using $\lambda_{p}=405$ nm and $\lambda_{s}=\lambda_{i}=810$ nm.}
\label{geoandangle}
\end{figure}

The phase matching process for SPDC determines the emission direction of the daughter photons. In general, 
perfect phase matching occurs when
\begin{equation}
 \vec{k}_{p}=\vec{k}_{s}+\vec{k}_{i},
\end{equation}
where 
\begin{equation}
 \vec{k}_{j} = \frac{n_{j}(\omega_{j}, \hat{s}_{j})\omega_{j}}{c} \hat{s}_{j},
\end{equation}
for $j=(p,s,i)$, where the angle- and frequency-dependent refractive index is given by $n_{j}(\omega_{j}, \hat{s}_{j})$, $\hat{s}_{j}$ is the 
propagation direction unit vector, and $c$ is the speed of light in vacuum. We follow the geometry, notations, and conventions in Refs. \cite{Beouff,Roberts}.
The crystal geometry and photon wavevectors and angles are shown in Fig. \ref{geoandangle}(a). A pump photon with wavevector $\vec{k_{p}}$ is incident on a nonlinear crystal and makes angles 
($\theta_{p},\phi_{p}$) with the optic axis of the crystal denoted ``OA-U" (``OA-B") for a uniaxial (biaxial) crystal, discussed in more detail below.
 The signal (idler) photon with wavevector $\vec{k_{s(i)}}$ is emitted at local angles ($\theta_{s(i)},\phi_{s(i)}$) with respect to $\vec{k_{p}}$. For perfect phase matching, $\phi_{s} = \phi_{i} + \pi$. The photons undergo refraction at the air-crystal interface and exit at exterior angles of $\theta_{s(i)}'$.

To determine the emission angles ($\theta_{s},\theta_{i},\phi_{s},\phi_{i}$) of the daughter photons, we choose frequencies ($\omega_{p},\omega_{s},\omega_{i}$) as well as the angles 
the pump wavevector makes with the  optic axis ($\theta_{p},\phi_{p}$). Using the
 Sellmeier equations for either BBO or BiBO, we solve for either set of parameters by finding a solution to the simultaneous equation \cite{Beouff}

\begin{align}
(\Delta k_{x})^{2}+(\Delta k_{y})^{2} + (\Delta k_{z})^{2} = 0, 
\label{totaldk}
\end{align}
and
\begin{equation}
 \Delta k_{z} = 0,
 \label{dkz}
\end{equation}
where, for example, $\Delta k_{z} = k_{p_{z}}-k_{s_{z}}-k_{i_{z}}$.
Evaluating Eqs. \ref{totaldk} and \ref{dkz} involves the refractive index for each photon propagating through the crystal, which depends on the polarization of the photons, and the propagation direction.

In a uniaxial crystal, which has a single axis of symmetry, an ordinary-polarized photon ($o$-polarized) is polarized perpendicular to the plane that contains its propagation wavevector and the optic axis of the crystal. This photon experiences a refractive index that does not change with the direction of propagation.  The extraordinary-polarized photon ($e$-polarized) is polarized in the plane of the optic 
axis and the propagation vector and experiences an angle-dependent refractive index. Type-I interactions, such as those considered here, include interactions where the pump is an $e$-polarized photon and the daughter photons
 are $o$-polarized, or vice versa.

A biaxial crystal, such as BiBO, has two optic axes and reduced symmetry. Polarized photons are neither $e$-polarized or $o$-polarized in the sense of a uniaxial crystal.
Depending on type of crystal and wavelength range, polarized photons may experience either an angle-dependent or angle-independent refractive index.
The photons are said to be either ``fast" or ``slow" instead of ``$e$" or ``$o$" where fast (slow) refers to having a smaller (larger) 
refractive index \cite{Roberts}.

Determining the fast (slow) refractive indices involves finding the length of the minor (major) axes of the optical indicatrix given by Fresnel's equation of wave normals given by
\begin{equation}
\frac{s_{x}^{2}}{n^{-2}(\omega,\hat{s})-n_{x}^{-2}}+\frac{s_{y}^{2}}{n^{-2}(\omega,\hat{s})-n_{y}^{-2}}+\frac{s_{z}^{2}}{n^{-2}(\omega,\hat{s})-n_{z}^{-2}}=0,
\label{wavenormals}
\end{equation}
where $n_{x},n_{y},n_{z}$ are the refractive indices in each principle direction of the crystal at a given vacuum frequency and $n(\omega,\hat{s})$ is the refractive index in a given direction 
with unit vector $\hat{s}$. For a negative biaxial crystal, such as BiBO, $n_{x}<n_{y}<n_{z}$. Solving Eq. \ref{wavenormals} for $n(\omega,\hat{s})$ using the approach described in Ref. \cite{Beouff}, we obtain two solutions, 
one for each polarization (fast and slow).
In BiBO we find that the pump photons experience the fast refractive index that is angle-independent for wavelengths in the UV and blue part of the spectrum. In contrast, the daughter photons in the
 red and NIR part of the spectrum experience the slow refractive index that is angle-dependent. 
In BBO, pump photons in the blue part of the spectrum travel as $e$-polarized photons, and experience an angle-dependent refractive index, while the signal and idler photons are $o$-polarized and experience an angle-independent refractive index. This effect is not due to the uniaxial versus biaxial nature of the crystals, but simply due to the angle dependence on the refractive index over certain wavelength ranges for each particular crystal.
Our analysis of the elliptical emission pattern for BiBO has not been noted or observed previously. This may be due to the fact that most previous quantum optics experiments have been conducted with BBO, which produces circular rings as observed and predicted below. 
This is not the case in BiBO due to the angle-dependent refractive index experienced by the daughter photons. In Fig. \ref{geoandangle}(b) we plot the external angle $\theta_{s}'$ as a function of its azimuthal angle $\phi_{s}$ for BiBO (blue, solid) and BBO (maroon, dashed). 
For BiBO, we observe that $\theta_{s}'$ varies with $\phi_{s}$ so that $\theta_{s}'$ has a larger value at $\phi_{s} = 0^{\circ}\hspace{2pt} \textrm{and} \hspace{2pt} 180^{\circ}$
than in the $\phi_{s} = 90^{\circ}\hspace{2pt} \textrm{and} \hspace{2pt}270^{\circ}$ directions. This leads to an elliptical emission pattern for this crystal. For BBO, $\theta_{s}'$
 is a constant as a function of $\phi_{s}$, which implies a circular emission pattern. In our analysis, we ignore birefringent walk-off of the beams due to its negligible contribution for the thin crystals considered here, as discussed in detail in the Appendix. 

\section{Experimental Results}
\begin{figure}[!h]
\begin{center}
\includegraphics[scale=0.35]{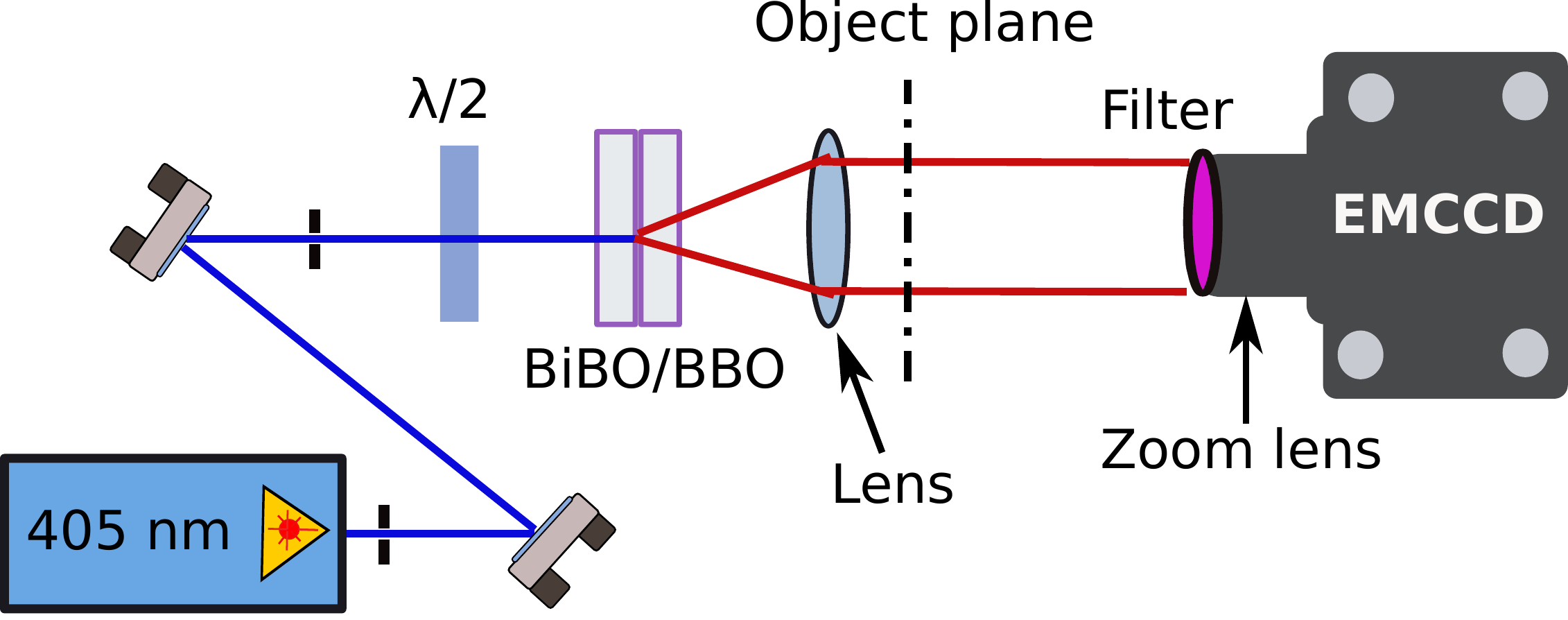}
\end{center}
\caption{Experimental setup. A 405 nm laser (Omicron, LDM405.120.CWA.L.WS, $<0.02$nm bandwidth FWHM) pumps either a BiBO or BBO crystal (Newlight Photonics). The down-converted light exits the crystal and propagates 
12.4 cm before it passes through a 40 cm focal length lens (Thorlabs LAC726B). This lens functions to direct the rays from the down-conversion ring to an object plane that we image onto the camera. We image the ring $\sim$ 10 cm after this lens by
 placing a zoom lens (Navitar Zoom 7000E) on an EMCCD (Andor $\textrm{iXon}^{EM}$) $\sim$ 1 m away from the lens. The magnification of this imaging system is 8.6.  Each pixel on the camera chip is 24 
$\mu$m $\times$ 24 $\mu$m,  the sensitive area of the chip is
 3 mm $\times$ 3 mm, and we cool the chip down to $-70^{\circ}\textrm{C}$ to reduce dark noise. We put a 10 nm bandpass optical filter (Andover 810FS10-50) before the camera 
lens to select only nearly degenerate wavelengths.}
\label{expsetup}
\end{figure}

We image the spatial intensity patterns from both BiBO and BBO using the experimental setup shown in Fig. \ref{expsetup}. A 405-nm-wavelength, continuous-wave laser pumps either a 
BBO or BiBO crystal.  Both types of crystals are thin (0.8 mm) in comparison to the Rayleigh length and compared to the transverse extent of the imaged rings - a requirement for polarization entanglement. As the crystals become thick, which-path information begins to degrade the entanglement because the rings become distinguishable.
 We test two different BiBO crystals each with a different set of 
crystal cut angles ($\theta_{p}, \phi_{p}$) designed to be phase matched for $\lambda_{p}=405$ nm and $\lambda_{s}=\lambda_{i}=810$ nm with an exterior opening angle of $\sim3^{\circ}$. These two crystal cut angles are the angles for which it is straight-forward to create an optical beam that propagates through the crystal with negligible walk-off. That is, $\phi_{p}$ is chosen so that the pump, signal, and idler photons essentially propagate as a slow or a fast wave. We can further tilt the crystal around $\theta_{p}$ to tune the opening angle. One crystal has cut angles of $\phi_{p} = 90^{\circ},\theta_{p} = 151.7^{\circ}$  while the other has 
$ \phi_{p} = 0^{\circ},\theta_{p} = 51^{\circ}$. The BBO crystal has a pump cut angle of $\theta_{p} = 29.3^{\circ}$. 

 Making precise measurements of the eccentricity requires having the well-calibrated imaging system, depicted in Fig. \ref{expsetup}. To achieve this, we perform the following procedure: first we remove the crystal 
and the 40 cm focal length lens, leaving only the laser, a set of apertures for alignment, steering mirrors, and camera-lens system. We then place a flat 
mirror against the camera
 lens to ensure the back-reflected light is going straight back through the apertures. We then place a target  in the object plane. The target is a flat 
piece of metal with a 20-mm-diameter ring scored in the surface of the metal, where the diameter tolerances is $\sim 25$ $\mu$m.  This machined ring has 
a hole in the center so that we can easily align it with the laser beam path and check its back reflections for tilt. We check the
 eccentricity of the target and make small adjustments to the camera's position and tilt until we minimize eccentricity. The eccentricity is caused primarily 
by any amount of tilt in the system, which arises from imperfect alignment. Astigmatism and coma also arise from an imperfectly aligned optical system, 
although these are negligible compared to the tilt for a well-aligned optical system. We then replace the crystal and check back reflections
 with a mirror on the camera lens. Finally, we add in the lens and ensure its alignment by checking back reflections. 
 
We collect images for each crystal and fit the observed emission pattern with an elliptical function in two transverse dimensions with a Gaussian 
profile in the longitudinal dimension.
 The free parameters of our model include the height of the Gaussian peak, 
background counts (offset of the Gaussian from 0), major/minor axes for ellipse, width of Gaussian and location of the center point. We calculate the 
eccentricity of the ellipse by
\begin{equation}
 \epsilon = \sqrt{1-\left(\frac{a}{b}\right)^{2}},
\end{equation}
where $b$ ($a$) is the major (minor) axis of the ellipse.

\begin{figure}[!h]
 \begin{center}
 \includegraphics[scale=0.3]{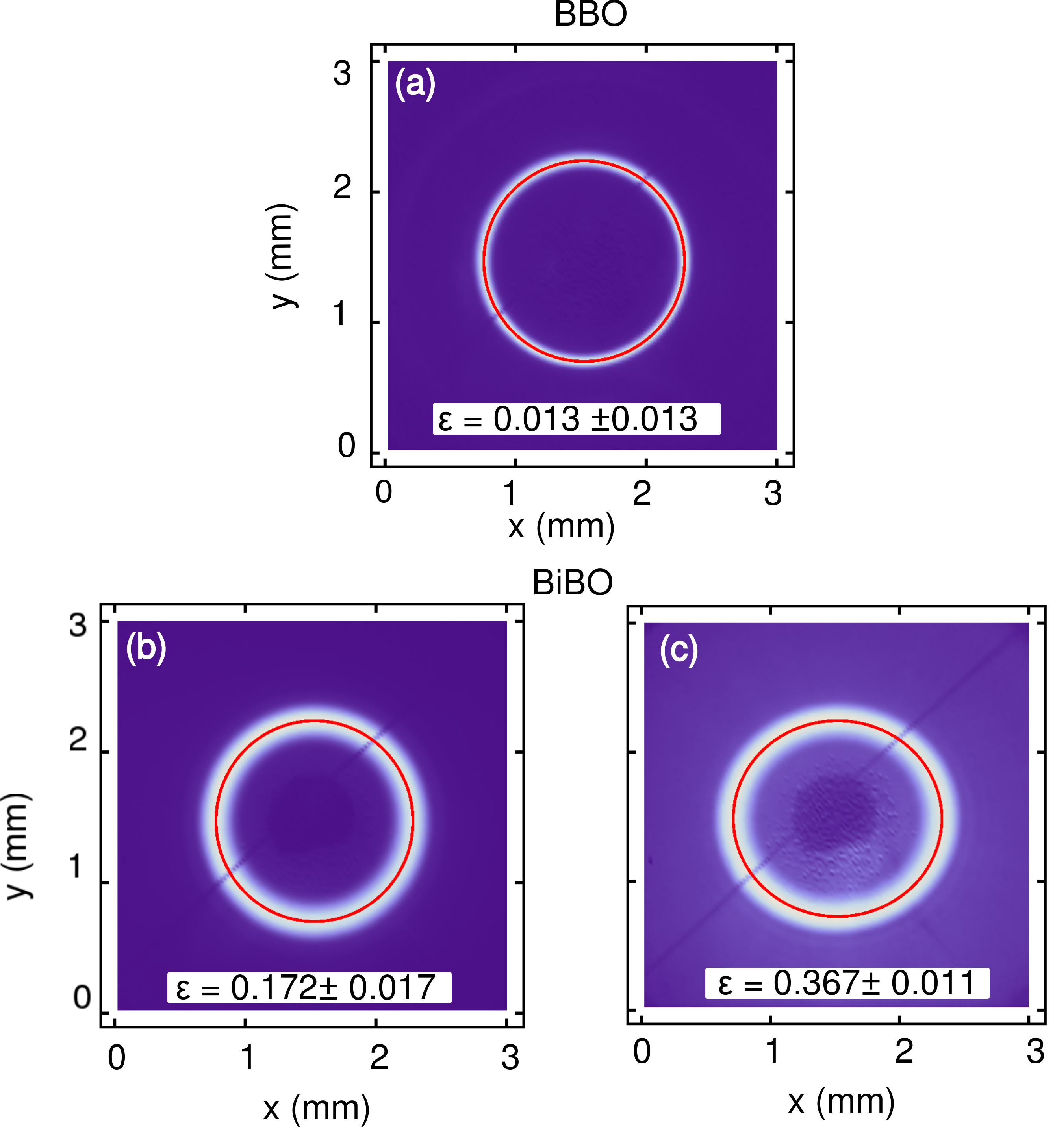}
\end{center}
\caption{SPDC emission pattern for BBO and BiBO. The emission pattern from the camera is plotted versus the two transverse dimensions ($x$ and $y$) where $y$ is in the direction perpendicular
to the optical table. The patterns from the camera have been scaled up by the magnification and converted from pixels into cm. The red, solid lines are ellipses with the major and minor axes taken from the fit parameters. 
(a) Down-conversion ring from BBO crystal is nearly circular  with an eccentricity of 0.013. 
(b) Down-conversion ring BiBO crystal cut at phase matching angle of ($\theta_{p}=151.7^{\circ},\phi_{p} = 90^{\circ}$) has a higher eccentricity with
the major axis in the x-direction. 
(c) Down-conversion ring in BiBO crystal cut at phase matching  angle of ($\theta_{p}=51^{\circ},\phi_{p} = 0^{\circ}$) has a large eccentricity.}
\label{data}
\end{figure}

 Figure \ref{data} shows the transverse spatial intensity pattern for down-conversion rings from both BBO and BiBO crystals. For BBO, the intensity pattern of the down-conversion ring is
essentially circular (Fig. \ref{data} (a)),  $\epsilon=0\pm$0.013 where the error is a combined statistical and systematic error of 0.013, which will be 
discussed below. Hence, our results indicate the pattern for BBO is circular to within our experimental uncertainties.
The range of opening angles that are phase matched for BBO is smaller than that for BiBO because the slope of the opening angle versus wavelength 
is larger in BBO due to the slope of the refractive index versus angle being steeper. This leads to a thinner down-conversion ring for BBO.
For BiBO cut at
phase matching angles $ \phi_{p} = 90^{\circ}, \theta_{p} = 151.7^{\circ}$, (Fig. \ref{data}, (b)), the eccentricity is greater than that for BBO, with 
$\epsilon = 0.172 \pm 0.019$. For BiBO, with $\phi_{p} = 0^{\circ}, \theta_{p} = 51^{\circ}$ (Fig. \ref{data} (c)), 
$\epsilon = 0.367 \pm 0.012$ and is easily seen by eye. 

  Table 1 gives the results of our observations and analysis.
We calculate the major/minor axes using the experimental values for $\theta_{s}'(\phi_{s}=0)/\theta_{s}'(\phi_{s}=90)$ and Eq. \ref{Eccentricity1}. Using the procedure outlined in Sec. 2, we calculate theoretical values for the eccentricity using an exterior angle for either the major or minor axis given 
in the first column. The exterior angle is what we measure experimentally and is simply the opening angles in the crystal propagated outside the crystal Snell's law. We use this set exterior angle to calculate the $\theta_{p}$ and $\phi_{p}$ that gives this value at either 
$\phi_{s} = 0$ or $\phi_{s} = 90^{\circ}$. In this case, instead of knowing $\theta_{p}$ and calculating $\theta_{s}$ and $\theta_{i}$, we know the
 latter and calculate the former.  Measuring the external emission angle is more straightforward than measuring the angle the pump beam makes with the crystal optic axis. Once we determine $\theta_{p}$, we use it to predict the emission angle around the entire ring.
  Using Snell's law, we propagate these angles outside the crystal and find the eccentricity through the relation,
\begin{equation}
\epsilon = \sqrt{1-\left(\frac{\tan[\theta_{s}'(\phi_{s}=0^{\circ})]}{\tan[\theta_{s}'(\phi_{s}=90^{\circ})]}\right)^{\pm2}},
\label{Eccentricity1}
\end{equation} 
where the $\pm$ accounts for the possibility that the axes at $\phi_{s}=0$ may be the major or the minor axis. We use Eq. \ref{Eccentricity1} to calculate both experimental and theoretical values for $\epsilon$. For the experimental values, we have measured both $\tan[\theta_{s}'(\phi_{s}=90^{\circ})]$ and $\tan[\theta_{s}'(\phi_{s}=0^{\circ})]$, while for the theoretical values, we use only one of these, either $\tan[\theta_{s}'(\phi_{s}=90^{\circ})]$ or $\tan[\theta_{s}'(\phi_{s}=0^{\circ})]$ and calculate the other by the procedure described above.
The error in our experiments includes a statistical error from the fitting process and a systematic error, which is most likely due to small astigmatism in the system.  We calculate the systematic error by taking repeated measurements at different times which requires realignment using the alignment procedure outlined above. 


\begin{table}
 \caption{Experimental Data}
 \begin{center}
\begin{tabular}[h,c]{||p{1.55cm}p{1.95cm}||p{1.15cm}|p{1.55cm}|l||}
\hline
&&{BBO}&{BiBO}&{BiBO}\\
&&&$\phi_{p}=90^{\circ}$&$\phi_{p}=0^{\circ}$\\
\hline
 Experiment&$\frac{\theta_{s}'(\phi_{s}=0)}{\theta_{s}'(\phi_{s}=90)}$ ($^{\circ}$) & $\frac{4.12358}{4.12382}$ &$\frac{4.0294}{4.10893}$&$\frac{4.05449}{4.31798}$ \\
 &$\epsilon$& 0.013 & 0.172& 0.360\\
\hline
Theory&$\epsilon$ &0&0.166& 0.361\\
\hline
Error&  &&&  \\
Statistical&&0.011& 0.001&0.001\\
Systematic &&0.007&0.019 &0.012 \ \\
\hline
\hline
\end{tabular}
\end{center}
\end{table}

\section{Theoretical Model}
  
In this section we outline a theoretical model for predicting the approximate eccentricity of the emission pattern in each crystal.  Because our emission angles are small ($\theta_{s},\theta_{i}<<1$), a small-angle approximation around the collinear case ($\theta_{s} =\theta_{i} =0$) is a natural method for obtaining an approximate analytic solution. 
This allows us to gain better understanding of the parameters that affect the eccentricity.
Starting from the phase matching equations in the $y$ and $z$ directions, we find that the phase matching occurs when,
\begin{equation}
 n_{s} \cos(\theta_{s}) +n_{i} \cos(\theta_{i}) -2n_{p}=0,
 \label{PMz}
\end{equation}
\begin{equation}
 n_{s} \sin(\theta_{s}) +n_{i} \sin(\theta_{i}) =0.
 \label{PMy}
\end{equation}

We expand $\sin(\theta_{j})$, $\cos(\theta_{j})$, $n_{s}$, and $n_{i}$  to second order in $\theta_{j}$. The approximate expressions of the refractive indices are given by,
\begin{align}
 n_{s} &\approx \tilde{n}_{s}+\frac{\partial\tilde{n}_{s}}{\partial \theta_{p}}\delta\theta_{p}+\frac{\partial\tilde{n}_{s}}{\partial \theta_{s}}\theta_{s}+\frac{1}{2}
\frac{\partial^{2}\tilde{n}_{s}}{\partial \theta_{p}^{2}}\delta\theta_{p}^{2}+\\ \nonumber
&\hspace{90pt}\frac{\partial^{2}\tilde{n}_{s}}{\partial\theta_{p}\partial\theta_{s}}\delta\theta_{p}
\theta_{s}+\frac{1}{2}\frac{\partial^{2}\tilde{n}_{s}}{\partial \theta_{s}^{2}}\theta_{s}^{2},
\label{nsapprox}
\end{align}
\begin{align}
 n_{i} &\approx  \tilde{n}_{s} +\frac{\partial\tilde{n}_{s}}{\partial \theta_{p}}\delta\theta_{p}-\frac{\partial\tilde{n}_{s}}{\partial \theta_{s}}\theta_{i}+\frac{1}{2}
\frac{\partial^{2}\tilde{n}_{s}}{\partial \theta_{p}^{2}}\delta\theta_{p}^{2}-\\ \nonumber
&\hspace{90pt}\frac{\partial^{2}\tilde{n}_{s}}{\partial\theta_{p}\partial\theta_{s}}\delta\theta_{p}
\theta_{i}+\frac{1}{2}\frac{\partial^{2}\tilde{n}_{s}}{\partial \theta_{s}^{2}}\theta_{i}^{2},
\label{niapprox}
\end{align}
where $\theta_{p}=\theta_{p_{0}}+\delta \theta_{p}$ is the pump phase matching angle, $\theta_{p_{0}}$ is the pump phase matching angle for the collinear 
degenerate case and $\delta \theta_{p}$ is a small variation around that angle. The notation $\tilde{n_{s}}$ describes the refractive index for the collinear degenerate case,
and $\tilde{n_{s}}$ has replaced $\tilde{n_{i}}$ everywhere because for the collinear degenerate case, they are equal. The change in refractive index with 
angle is calculated from the expression for the frequency and directional dependent refractive index. These quantities indicate how quickly the refractive 
index changes with the various angles.
Inserting Eqs. \ref{nsapprox} and \ref{niapprox} and the trigonometric functions into Eqs. \ref{PMz} and \ref{PMy}, we solve for angles $\theta_{s}$ and
 $\theta_{i}$. We are interested in solving for $\delta\theta_{s}$ and $\delta\theta_{i}$ which are small changes in the emissions angles around the values 
of $\theta_{s}(\phi_{s}=90^{\circ})$ and $\theta_{s}(\phi_{s}=0^{\circ})$, which are the angles corresponding to the directions of the major and minor axes.
 We use the expression
\begin{equation}
 \theta_{s,(i)}=\theta_{s}(\phi_{s}=90^{\circ})+\delta\theta_{s,(i)},
\end{equation}
to solve for $\delta\theta{s}(\phi_{s}=0^{\circ})$ and  $\delta\theta_{s}(\phi_{s}=180^{\circ})$. We calculate the eccentricity inside the crystal to be
\begin{equation}
 \epsilon=\sqrt{1-\left(\frac{\delta\theta_{s}(\phi_{s}=0^{\circ})+\delta\theta_{s}(\phi_{s}=180^{\circ})}{2\theta_{s}(\phi_{s}=90^{\circ})}\right)^{2}}.
 \label{eccen1}
\end{equation}
The eccentricity arrises from the photons experiencing the angle-dependent refractive index, so the eccentricity in the emission pattern is only due to the eccentricity inside the 
crystal. Once the photons propagate into free-space, they experience no angle-dependence and simply follow the trajectories of the external angles dictated by Snell's law. Therefore, 
the eccentricity inside the crystal is the only contribution to the eccentricity experimentally measured.

\begin{figure}
\begin{center}
 \includegraphics[scale=0.35]{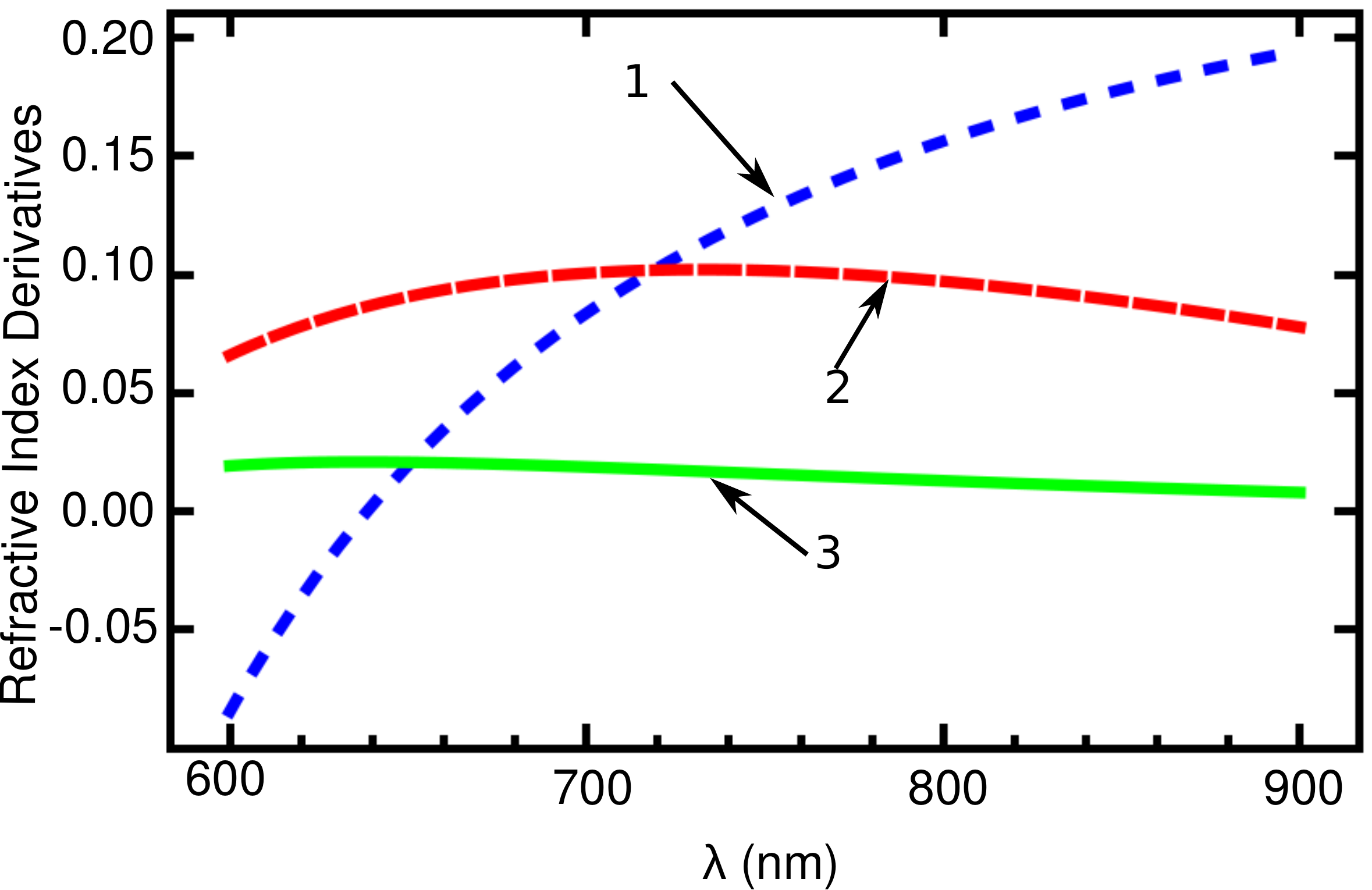}
\end{center}
\caption{Plot of each term in Eq.\ref{approxe}  versus wavelength, ``1" (blue, short-dash) $(\partial^{2}\tilde{n_{s}}/\partial\theta_{s}^{2})^{\phi_{s}=0}$
``2" (red, long-dash) $(\partial^{2}\tilde{n_{s}}/\partial\theta_{s}^{2})^{\phi_{s}=90^{\circ}}$, ``3" (green, solid) $2/\tilde{n_{s}}*[(\partial\tilde{n_{s}}/\partial\theta_{s})^{\phi_{s}=0}]^{2}$. 
For this plot we use $\phi_{p} = 90^{\circ},\theta_{p} = 152.077^{\circ}$.
}
\label{threederivatives}
\end{figure}

\begin{figure}
\begin{center}
 \includegraphics[scale=0.35]{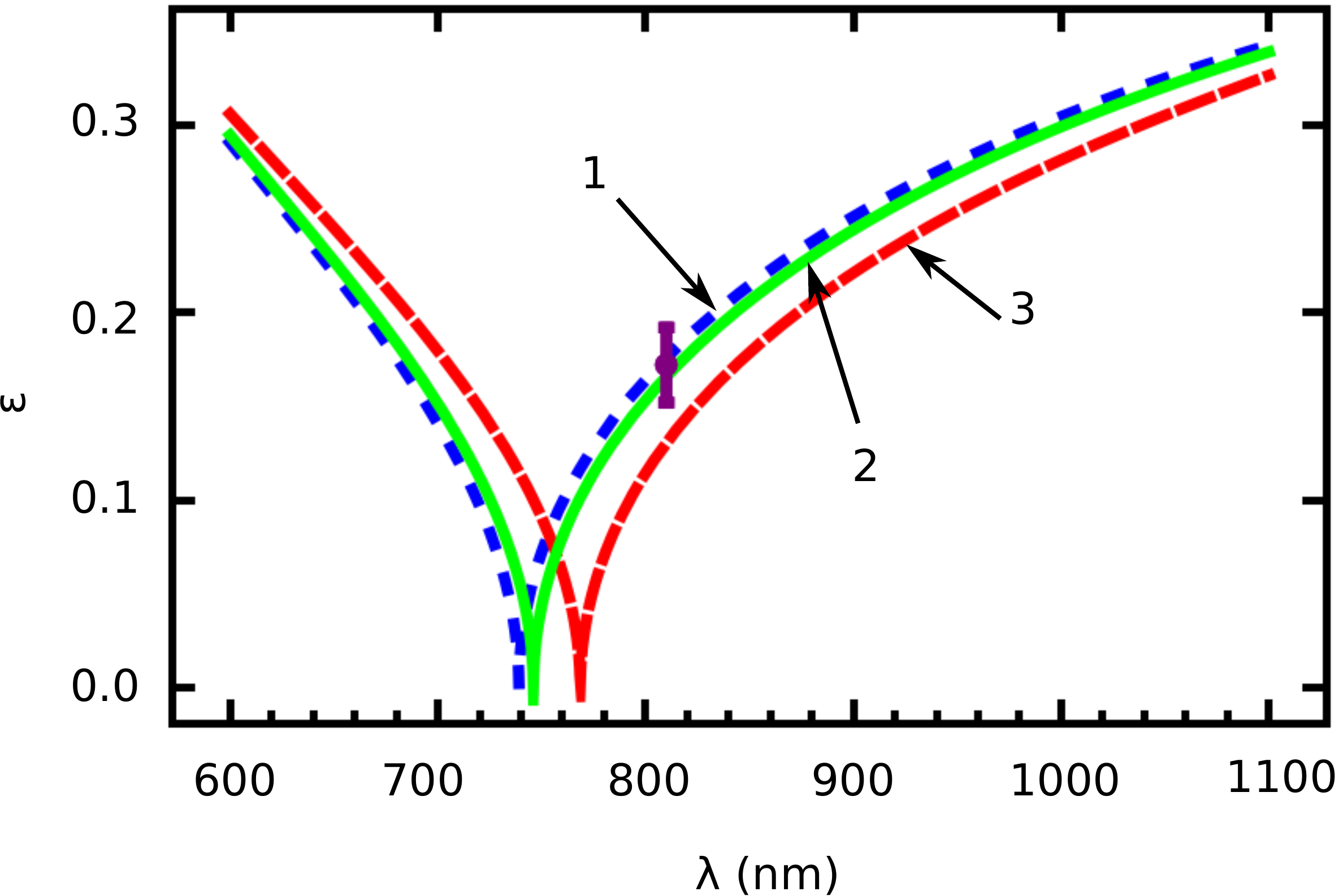}
\end{center}
\caption{Eccentricity versus wavelength for three different values of $ \theta_{p}$, ``1" (blue, short-dash) $\theta_{p}=152.071^{\circ}$
``2" (green,solid) $\theta_{p}=151.378^{\circ}$, ``3" (red, long-dash) $\theta_{p}=149.21^{\circ}$. 
For all three plots there is a wavelength for which the eccentricity of the down-conversion ring is minimized. 
We show our experimental data for a pump wavelength of 405 nm and a degenerate down-converted wavelength of 810 nm (purple) with associated error bars.}
\label{magicw}
\end{figure}

Using Eq. \ref{eccen1}, we determine the predicted value of the eccentricity using the measured values for $\delta\theta_{p}$. 
These are shown in the final columns of Table 1. For small values of $\delta\theta_{p}$, we also express the eccentricity in terms of the derivatives of refractive indices as,
\begin{equation}
\epsilon \approx \bigg(\frac{\partial^{2}\tilde{n_{s}}}{\partial\theta_{s}^{2}}\bigg)^{\phi_{s}=0}-
\bigg(\frac{\partial^{2}\tilde{n_{s}}}{\partial\theta_{s}^{2}}\bigg)^{\phi_{s}=90^{\circ}}-
\frac{2}{\tilde{n_{s}}}\bigg[\bigg(\frac{\partial\tilde{n_{s}}}{\partial\theta_{s}}\bigg)^{\phi_{s}=0}\bigg]^{2},
\label{approxe}
\end{equation}
where the superscripts of $\phi_{s}$ denote the azimuthal angle at which the terms are evaluated. We plot the three terms in this approximation in Fig. \ref{threederivatives}.
For this plot, the difference between the curves ``1'' and ``2'' is the dominant contribution to the eccentricity. Where these curves intersect is where the eccentricity is at a minimum in
Fig. \ref{magicw}.
This expression allows us to determine why eccentricity is larger for certain crystal cuts than others. We examine the relative size of each of the three terms in Eq. \ref{approxe} for
the case where $\phi_{p}=0$ and $\phi_{p}=90^{\circ}$. We find that the difference of the first two terms in Eq. \ref{approxe} for $\phi_{p}=0$ is very large in comparison to the difference
of the first two terms for the $\phi_{p}=90^{\circ}$ case. This ultimately leads to the larger eccentricity we observe for the $\phi_{p}=0$ crystal cut in BiBO.

Using this method, we also calculate the value of the degenerate wavelength ($\lambda_{s}=\lambda_{i}$) that minimizes the eccentricity inside the crystal. Again, minimizing the internal eccentricity will minimize the eccentricity measured in the external emission pattern. 
Each of the parameters in Eq. \ref{eccen1} is wavelength-dependent because the refractive indices for the daughter photons in BiBO are angle-dependent. 
We find a minimum wavelength by performing a root-finding routine on Eq. \ref{eccen1} with wavelength as the independent parameter. We also find the 
eccentricity as a function of wavelength plotted in Fig. \ref{magicw} for three different values of
$\theta_{p}$. This plot shows that there is a wavelength for each pump tilt angle for which the eccentricity can be minimized. According to the plot, the eccentricity becomes large away from the 710-780 nm range. For our degenerate wavelength of 810 nm, this figure agrees with the value of the eccentricity we observe. This plot indicates which wavelengths are best at minimizing eccentricity in a given crystal cut for Type-I degenerate SPDC. 

\section{Impact on entanglement and count rate}
One potential drawback of an elliptical emission pattern is that it may decrease the entanglement quality and/or the entangled photon count rate. Because the rings are elliptical, 
they no longer perfectly spatially overlap around the entire ring, leading to potential degradation in entanglement-purity, and limiting the ability to multiplex many channels
 around the down-conversion ring. Lack of spatial overlap is the most obvious way in which which-path information is revealed. The elliptical shape may also cause a change in the joint 
spectrum for the photons, which leads to a degradation in entanglement purity. We note that spatial overlap can be corrected for by collecting light in between the ring centroids
 at the cost of count rate. However, at this location, the spectra from the two rings is different, leading to decreased entanglement purity. We study each of these effects in 
different ways where we assume that the biphotons are coupled into single mode fibers. 

To estimate whether the entanglement quality is
 degraded, we determine the spectrum of the down-converted light for each crystal for single mode fiber collection. Using single mode fibers is important because collection of a 
single spatial mode for either the signal or idler photon means that the twin photon should also be projected onto a single spatial mode. The entanglement purity can then be determined 
from the overlap integral of the spectral intensity of these two single modes.
Calculating the spectra from the midpoint between the rings at two locations and computing the overlap integral provides a
 measure of distinguishability. A smaller overlap integral represents lower entanglement quality because the difference in spectra between the crystals means it is possible to 
distinguish which crystal
 the photons came from. 

\begin{figure}[!h]
\begin{center}
 \includegraphics[scale=0.35]{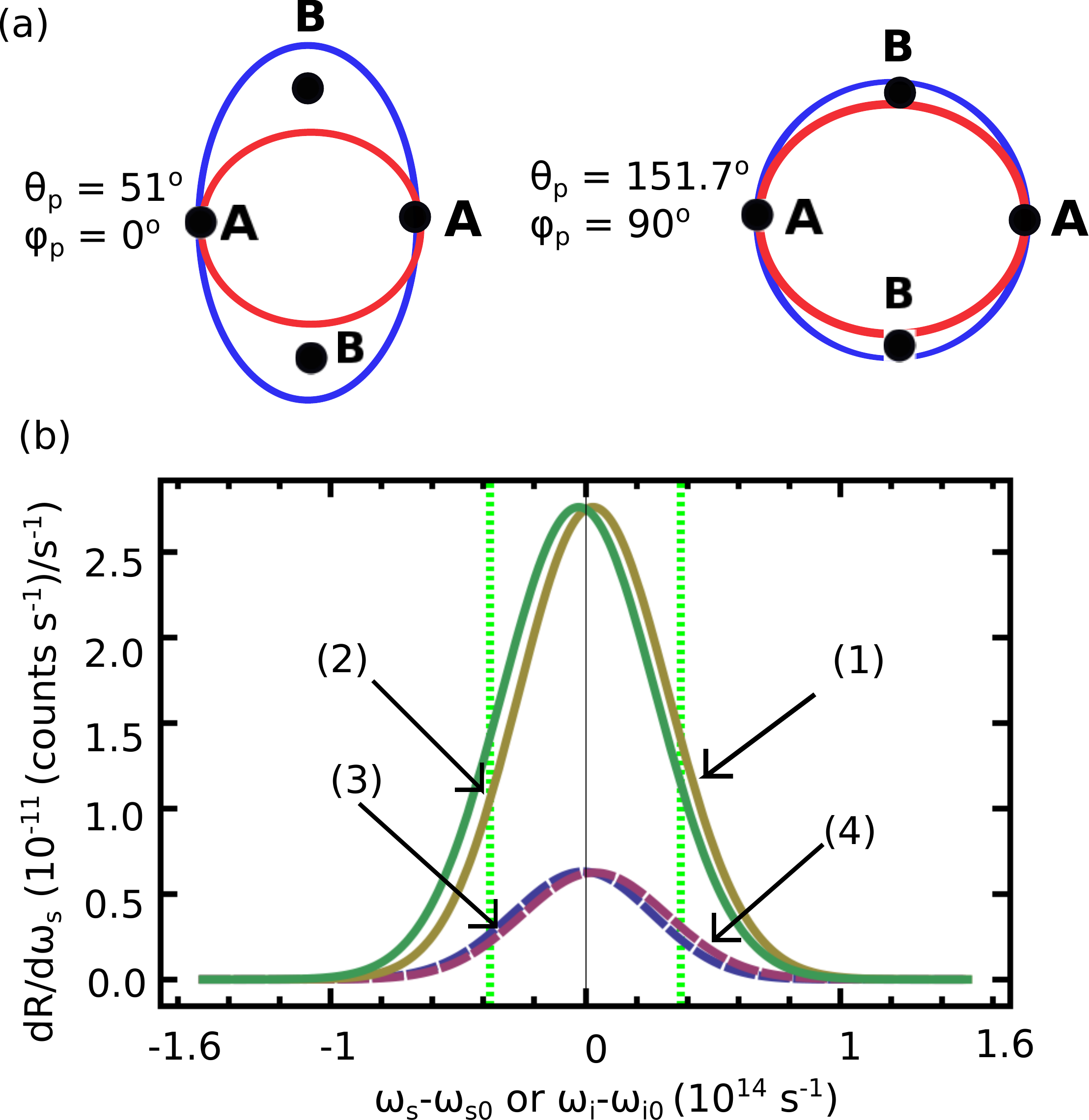}
\end{center}
\caption{(a) Points A and B represent the points of maximal and minimal overlap respectively. Red inner ring lines and blue outer ring lines illustrate the center of the emission patterns from each crystal 
in the crystal pair, where we have ignored the finite width of the actual ring. The eccentricity of these rings has been exaggerated for illustration purposes. Left figure depicts crystal cut at $\phi_{p}=0, \theta_{p} = 51^{\circ}$ and right figure depicts crystal cut at  
$\phi_{p}=90^{\circ}, \theta_{p} = 151.7^{\circ}$(b) Joint spectra for each crystal cut at location B for each crystal cut. The $y$ axis on the plot is 
the differential rate per frequency and the $x$ axis is an angular frequency shift around the degenerate frequency. Solid lines (green (2) and gold (1)) represent the joint 
spectral for each crystal when the crystals are cut at $\phi_{p}=90^{\circ}, \theta_{p} = 151.7^{\circ}$ while the dashed lines (blue (3) and purple (4)) represent the joint 
spectra from each crystal when the crystals are cut at $\phi_{p}=0, \theta_{p} = 51^{\circ}$. The pump wavelength for this simulation is 405 nm and the degenerate wavelengths 810 nm. The vertical dashed green lines in the plot indicate a 20 nm bandwidth for reference.}
\label{final}
\end{figure}

Using the formalism developed in our previous work \cite{Guilbert}, we determine both the joint spectrum for the biphoton wavefunction as well as the singles spectrum for each signal 
and idler, where we assume single mode fiber collection. Figure \ref{final} (b) shows the joint spectrum at location B in Fig. \ref{final} (a). Here the overlap is minimal for each crystal in the pair when the crystals are cut at 
 $\phi_{p}=90^{\circ}, \theta_{p} = 151.7^{\circ}$ (solid lines) and $\phi_{p}=0, \theta_{p} = 51^{\circ}$ (dashed lines). In these calculations we choose one of these points,
 point A (Fig. \ref{final} (a)) to have maximal overlap between the two rings, leaving point B to have minimal overlap. We choose the minimal-overlapped case because the entanglement purity is lowest due to the mismatched overlap. 
 For the minimal overlap case, we choose the collection mode to be in 
between the two rings so that we collect equal count rates from each and so that the spectra from the two will be most similar. We find that, for both BiBO crystal cuts, the spectra are similar for the two crystals,
 and the entanglement quality is not substantially degraded. The spectra at point A are identical, while at point B we find the overlap integral is 99.4-99.97$\%$ for each of the 
two crystal cuts. The overlap integral for two entangled states is a measure of the entanglement purity of the states involved. 

Even though the eccentricity of the ring does not affect the entanglement purity, it does affect the count rates. Integrating the spectra in Fig. \ref{final} (b) over $\Delta \omega$ gives the total joint count rate. For the
 crystal cut at $\phi_{p}=0, \theta_{p} = 51^{\circ}$, the spectra has a much lower amplitude and integrating over any band of frequencies gives a lower count rate compared to the 
crystal cut at $\phi_{p}=90^{\circ}, \theta_{p} = 151.7^{\circ}$. Higher eccentricity causes the rings to be more spatially separated in the former case, resulting in the 
lower joint count rate. Additionally, the crystal cut for this high eccentricity ($\phi_{p}=0, \theta_{p} = 51^{\circ}$) has a lower effective nonlinearity ($d_{\rm{eff}}$) resulting in an even lower count rate so that this crystal cut is clearly not optimal for high-brightness applications. This may not be the case for other crystals however. For SPDC applications in BiBO that require both a high count rate as well as high purity entanglement, we find that the eccentricity does not decrease the entanglement purity significantly, however, the count rate is lower for collection of the midpoints between emissions patterns with greater eccentricity. Applications aiming for ultra-high entanglement purity ($> 99.99\%$) are limited to the locations around the ring that can be made to overlap completely. 
\section{Conclusions}
In conclusion, we show both experimentally and theoretically that the emission patterns from down-conversion in BiBO have an elliptical shape, while
the pattern from BBO has a circular profile. We show that this difference in shape of the down-conversion rings
depends on whether or not the daughter photons experience an angle-independent or angle-dependent refractive index. Although, in hindsight, the results we present may seems obvious, this is to our knowledge, the first time this effect has ever been characterized and its potential impact discussed. We
also present a theoretical method for calculating the eccentricity of the down-conversion ring for biaxial crystals find that there is an optimal wavelength
for which the eccentricity is a minimum and close to zero. We demonstrate that the elliptical nature of the rings does not reduce entanglement purity, but reduces joint counts rates significantly for down-conversion patterns with a larger eccentricity.
\section{Acknowledgments}
The authors gratefully acknowledge the financial support of the DARPA InPho program and the Office of Naval Research MURI on Fundamental Research on Wavelength-Agile High-Rate Quantum Key Distribution (QKD) in a Marine Environment, award $\#$N00014-13-0627.  
\section*{Appendix}
Birefringent walk-off occurs in a medium when, for a particular polarization, the momentum vector and the Poynting vector separate from each other. Walk-off 
occurs in both uniaxial and
biaxial crystals, although calculating the walk-off angle in the biaxial case is more challenging due to the reduced symmetry of the crystal. We calculate the 
walk-off angles using the method in Ref. \cite{Roberts}, 
and find the contribution to the eccentricity from walk-off is negligible, due to the fact that the crystals considered here are so thin. 
The crystal thinness means that photons that are walking off different amounts in different directions will not cause a large change in eccentricity because 
their propagation distance to the crystal face is quite small.

In BiBO, the daughter photons experience an angle-dependent refractive index, and will therefore have Poynting vector walk-off. Using Eqs.~(20)-(22) in Ref. \cite{Roberts} we determine
the Poynting unit vector $\hat{N}$ and unit propagation vector $\hat{k}$. We use these to determine the walk-off angle as a function of azimuthal angle $\phi_{s}$ around the original pump 
direction. We find walk-off angles for the $\phi_{p} = 90^{\circ}, \theta_{p} = 151.56^{\circ}$ crystal cut at $\phi_{s} = 0$ and $\phi_{s}=180^{\circ}$ are
 $3.19^{\circ}$ and $3.51^{\circ}$, respectively. The walk-off angles for $\phi_{s} = 90^{\circ}$ and $\phi_{s} = 270^{\circ}$  are both $3.36^{\circ}$.
We map the Poynting vectors and momentum vectors onto the 
crystal exit face and compare their eccentricity. We find that the Poynting vectors map a ring with eccentricity of 0.1680, while the momentum 
vectors map a ring with eccentricity 0.1685, which is a 0.3$\%$ difference.

In free space, the Poynting vector and momentum vector must be parallel, so the trajectories of these vectors to the image plane are identical. As we have shown in the previous 
paragraph, the eccentricity of the ring as it is about to exit the crystal is negligible. 
The photon path outside the crystal is much larger than the one inside the crystal and as the Poynting vector and momentum vector are parallel in free-space, the dominant 
contribution to the eccentricity comes from the paths taken outside the crystal. Essentially, the overall impact of the walk-off from intracrystal angles is very small due 
to the fact that the crystal length is much smaller than the crystal - lens distance.







\bigskip


\begin{thebibliography}{25}
\bibitem{BB84} C. H. Bennett and G. Brassard, ``Quantum Cryptography: Public key distribution and coin tossing,'' in Proceedings of the IEEE International Conference on Computers, Systems, and Signal Processing, Bangalore, 175 (1984)
\bibitem{Ekert} A. Ekert, ``Quantum cryptography based on Bell's theorem,'' Phys. Rev. Lett. \textbf{67}, 661-663 (1991).
\bibitem{Serigenko} A. V. Sergienko, M.  Atat\"ure, Z.  Walton, G. Jaeger, B. E. A. Saleh, and M. C. Teich, ``Quantum cryptography using femtosecond-pulsed parametric down-conversion," Phys. Rev. A \textbf{60}, R2622--R2625 (1999).
\bibitem{Zeilenger} M. Giustina, S. Ramelow, B. Wittmann, J. Kofler, J. Beyer, A. Lita, B. Calkins, T. Gerrits, S. W. Man, R. Ursin, A. Zeilinger, ``Bell Violation using entangled photons without the fair-sampling assumption," Nat. \textbf{497}, 227 (2013).
\bibitem{bradprl} B. G. Christensen, K. T. McCusker, J. B. Altepeter, B. Calkins, T. Gerrits, A. E. Lita, A. Miller, L. K. Shalm, Y. Zhang, S. W. Nam, N. Brunner, C. C. W. Lim, N. Gisin, P. G, Kwiat, 
Detection-Loophole-Free Test of Quantum Nonlocality, and Applications," Phys. Rev. Lett. \textbf{111}, 130406 (2013).
\bibitem{Kim} O. Kwon, Y-W. Cho, and Y-H. Kim, ``Single-mode coupling efficiencies of type-II spontaneous parametric down-conversion: Collinear, noncollinear, and beamlike phase matching," Opt. Express   \textbf{78}, 053825 (2008).
\bibitem{Wong} M. Fiorentino, C. E. Kuklewicz, F. N. C. Wong, ``Source of polarization entanglement in a single periodically poled KTiOP$\textrm{O}_{4}$ crystal with overlapping emission cones," Phys. Rev. A \textbf{13}, 127-135 (2005).
\bibitem{Kwiat95} P. G. Kwiat, K. Mattle, H. Weinfurter, and A. Zeilinger, ``New High-Intensity Source of Polarization-Entangled Photon Pairs,''
 Phys. Rev. Lett.  \textbf{75}, 4337-4341 (1995).
\bibitem{Kwiat99} P. G. Kwiat, E. Waks, A.G. White, I. Applebaum, and P. Eberhard, ``Ultrabright Source of Polarization-entangled Photons,'' 
Phys. Rev. A  \textbf{60}, R773-R776 (1999).
\bibitem{Kwiat05} J. Altepeter , E. Jeffrey, and P. Kwiat, ``Phase-compensated ultra-bright source of entangled photons," Opt. Express  \textbf{13}, 8951-8959 (2005).
\bibitem{Kwiat09} R. Rangarajan, M. Goggin, and P. Kwiat, ``Optimizing type-I polarization-entangled photons,'' Opt. Express  \textbf{17}, 18920-18932 (2009).
\bibitem{Hellwig} H. Hellwig, J. Liebertz, and L. Bohaty, ``Exceptional large nonlinear optical coefficients in the monoclinic 
bismuth borate $\textrm{BiB}_{3}\textrm{O}_{6}$ (BIBO)," Solid State Commun. \textbf{109}, 249-251 (1999).
\bibitem{Ghotbi} M. Ghotbi and M Ebrahim-Zadeh, ``Optical second harmonic generation properties of $\textrm{BiB}_{3}\textrm{O}_{6}$,'' Opt. Express  \textbf{12}, 6002-6018 (2004).
\bibitem{Beouff} N. Beouff, D. Branning, I. Chaperot, E. Dauler, S. Guerin, G. Jaeger, A. Muller and A. Migdall, ``Calculating characteristics of noncollinear
phase matching in uniaxial and biaxial crystals,'' Opt. Eng. \textbf{39}, 4, 1016-1024 (2000).
\bibitem{Roberts} D. Roberts, ``Simplified Characterization of Uniaxial and Biaxial Nonlinear Optical Crystals: A Plea for Standardization of Nomenclature 
and Conventions,''  IEEE J. of Quant. Elec. \textbf{28}, 2057-2071 (1992).
\bibitem{Ling} A. Ling, A. Lamas-Linares, and C. Kurtsiefer, ``Absolute emission rates of spontaneous parametric down-conversion into single transverse Gaussian modes," Phys. Rev. A \textbf{77}, 043834 (2008).
\bibitem{Guilbert} H. E. Guilbert and D. J. Gauthier, ``Enhancing Heralding Efficiency and Biphoton Rate in Type-I Spontaneous Parametric Down-Conversion," IEEE. J. Sel. Topics Quantum Electon. \textbf{21}, 6400610 (2015).
\end{thebibliography}

\end{document}